# Lightweight Intrusion Detection System Using a Hybrid CNN and ConvNeXt-Tiny Model for Internet of Things Networks

Fatemeh Roshanzadeh[1,2], Hamid Barati*[1,2], Ali Barati[1,2]

roshanzadeh@iau.ir, hamid.barati@iau.ac.ir, alibarati@iau.ac.ir

1. Department of Computer Engineering, Dezful Branch, Islamic Azad University, Dezful, Iran
2. 2. Institute of Artificial Intelligence and Social and Advanced Technologies, Dez.C., Islamic Azad University, Dezful, Iran

**Abstract**

The rapid expansion of Internet of Things (IoT) systems across various domains such as industry, smart cities, healthcare, manufacturing, and government services has led to a significant increase in security risks, threatening data integrity, confidentiality, and availability. Consequently, ensuring the security and resilience of IoT systems has become a critical requirement. In this paper, we propose a lightweight and efficient intrusion detection system (IDS) for IoT environments, leveraging a hybrid model of CNN and ConvNeXt-Tiny. The proposed method is designed to detect and classify different types of network attacks, particularly botnet and malicious traffic, while the lightweight ConvNeXt-Tiny architecture enables effective deployment in resource-constrained devices and networks. A real-world dataset comprising both benign and malicious network packets collected from practical IoT scenarios was employed in the experiments. The results demonstrate that the proposed method achieves high accuracy while significantly reducing training and inference time compared to more complex models. Specifically, the system attained 99.63% accuracy in the testing phase, 99.67% accuracy in the training phase, and an error rate of 0.0107 across eight classes, while maintaining short response times and low resource consumption. These findings highlight the effectiveness of the proposed method in detecting and classifying attacks in real-world IoT environments, indicating that the lightweight architecture can serve as a practical alternative to complex and resource-intensive approaches in IoT network security.

## 1. Introduction

With the rapid advancement of digital technologies, the Internet of Things (IoT) has emerged as one of the key pillars of the Fourth Industrial Revolution, driving transformative changes across diverse domains. Recent statistics indicate that the number of active IoT devices worldwide has surpassed 26 billion and is expected to exceed 75 billion by 2025 [1]. IoT consists of a collection of intelligent physical objects interconnected through sensors, software, and other technologies, enabling them to exchange data with other devices and systems over the Internet [2,3].

As IoT infrastructures continue to expand, challenges such as data security, privacy preservation, and system reliability have gained increasing importance [4]. To address these challenges, researchers have explored approaches such as integrating blockchain technology into Industrial IoT (IIoT) networks, developing intrusion detection systems, and applying transfer learning algorithms [5,6]. Nevertheless, the demand for comprehensive, intelligent, and scalable models capable of real-time analysis of complex data in distributed environments remains pressing.

One of the emerging approaches to addressing security challenges in IoT environments is the use of game theory, which can contribute to energy optimization and enhanced security, particularly in applications such as smart campuses and industrial networks [3]. The Industrial Internet of Things (IIoT) has become a critical component in industries such as manufacturing, healthcare, energy, and intelligent transportation. However, the growing complexity and scale of IIoT environments expose these systems to threats such as botnet attacks, ransomware, and intrusions exploiting system vulnerabilities [7,8].

To counter these threats, artificial intelligence and machine learning (ML) have emerged as effective tools for analyzing data patterns and classifying threats [9–11], with deep learning being particularly highlighted for IoT security [10]. However, the inherent limitations of IoT devices, such as limited processing resources and low energy consumption, pose challenges to the implementation of these algorithms. In addition, issues such as model obsolescence over time and the occurrence of data bias affect the accuracy and reliability of the models [12–14]. Despite these limitations, numerous studies have demonstrated that deep learning–based approaches and hybrid cryptographic techniques can significantly enhance security and privacy in IIoT devices [9,13,14]. Therefore, the development of intelligent, lightweight, and attack-resilient learning frameworks will play a pivotal role in shaping the future of IoT ecosystems.

The proposed method in this paper introduces an innovative framework aimed at enhancing the accuracy, efficiency, and reliability of intrusion detection systems in Internet of Things (IoT) environments. With the growing prevalence of cyber threats, particularly sophisticated and targeted botnet attacks, the need for lightweight, scalable, and intelligent solutions to effectively counter these threats has become increasingly critical. Since IoT devices are typically constrained by limited computational resources, traditional security approaches are often ineffective in such contexts. Therefore, designing optimized architectures capable of processing data rapidly and identifying threats in real time is of vital importance.

Botnet attacks represent one of the most challenging threats to IoT infrastructures. By leveraging a network of compromised devices, botnets can perform data theft, launch denial-of-service (DDoS) attacks, and propagate malware. The complex structure, dynamic behavior of botnet networks, and the high volume of traffic in IoT environments make timely detection of such attacks particularly difficult. In response to this challenge, this paper proposes an efficient method based on a hybrid CNN and ConvNeXt-Tiny model, designed to build a lightweight yet accurate intrusion detection system for IoT devices.

By exploiting the deep feature extraction capability of CNNs and the superior generalization ability of the ConvNeXt-Tiny model, the proposed method demonstrates strong performance in analyzing network traffic data. Furthermore, the use of the well-established N-BaIoT dataset for training and evaluation ensures the empirical validity of the findings. Experimental results indicate that the proposed method outperforms existing advanced models in terms of accuracy and efficiency. This achievement can be considered a significant step toward strengthening cybersecurity in IoT networks.

The main contributions of this paper are summarized as follows:
- Development of a lightweight and efficient intrusion detection system based on a hybrid CNN and ConvNeXt-Tiny model, which offers high adaptability and scalability while effectively detecting and classifying various threats and attacks in IoT networks. The proposed method leverages the deep feature extraction capability of CNNs along with the optimized and lightweight architecture of ConvNeXt-Tiny to achieve optimal performance in resource-constrained IoT environments.
- Performance evaluation through a multi-class classification approach that distinguishes benign traffic from different types of attacks within IoT networks.
- Training and testing of the proposed method using the well-established N-BaIoT dataset, which contains real-world samples of botnet attacks.
- Comprehensive comparison of the proposed method with state-of-the-art techniques based on various metrics such as accuracy, efficiency, and processing speed, demonstrating significant improvements in intrusion detection performance for IoT networks.

The structure of this paper is organized as follows: Section 2 provides a comprehensive review of the existing literature on attack detection methods and identifies the common challenges associated with these approaches. Section 3 briefly introduces the fundamental concepts of Convolutional Neural Networks (CNN) and the ConvNeXt-Tiny model utilized in the proposed method. In Section 4, the proposed lightweight intrusion detection system, based on the hybrid architecture of CNN and ConvNeXt-Tiny for IoT networks, is presented and described in detail. Section 5 outlines the experimental design and data analysis of the proposed method, along with a comparison against state-of-the-art approaches. Finally, Section 6 summarizes the study's findings and highlights potential directions for future research.

## 2. Related Work

This section provides a systematic review of the existing literature on intrusion detection models based on various techniques.

Bakro et al. (2023) proposed an accurate and efficient intrusion detection system for cloud environments, leveraging a combination of intelligent feature selection and a weighted hybrid classifier. This system employs LSTM, SVM, XGBoost, and FLN algorithms, with weight optimization performed using the CSA method. It was evaluated on the NSL-KDD, Kyoto, and CSE-CIC-IDS-2018 datasets, achieving 99.99% accuracy, 99.87% precision, and a 99.91% F1-score. Its main advantage lies in its exceptionally high accuracy and low false alarm rate for scalable cloud network security [15].

Wang et al. (2023) designed a lightweight intrusion detection system named DL-BiLSTM for IoT, which combines DNN and BiLSTM models to extract features precisely. The system utilizes IPCA for dimensionality reduction, Optuna for optimization, and dynamic quantization, achieving high performance in resource-constrained environments. Evaluation on the CIC IDS2017, N-BaIoT, and CICIoT2023 datasets demonstrated an accuracy of 99.49%, outperforming existing methods. Its main advantage is the combination of high accuracy and lightweight design, making it suitable for scalable IoT applications [16].

Zhao et al. (2023) introduced a lightweight intrusion detection model named ConvNeXt-Sf for IoT environments within a cloud-fog framework, optimizing the model architecture for network data by combining ConvNeXt with ShuffleNet V2. This system was evaluated on the BoT-IoT and TON-IoT datasets and uses only 1.25% of the parameters of the original ConvNeXt model, significantly reducing training and inference time while increasing accuracy by 6.18%. The main advantage of this approach is the creation of a fast, accurate, and lightweight system suitable for fog nodes in IoT networks [17].

Bella et al. (2024) proposed an intrusion detection system for IoT named DNDF-IDS, which combines neural networks with an enhanced decision forest to achieve precise anomaly detection. By leveraging various feature selection techniques and evaluating on the NSL-KDD, CICIDS2017, and UNSW-NB15 datasets, the system achieved accuracy ranging from 94.09% to 98.84% with a prediction time of only 0.1 milliseconds. Its main advantage lies in high accuracy coupled with computational efficiency in resource-constrained IoT environments [18].

Chintapalli et al. (2024) introduced an effective IoT intrusion detection system called OOA-modified Bi-LSTM, which integrates the Osprey optimization algorithm with a modified Bi-LSTM network, achieving high accuracy in cyberattack detection. By selecting effective features and employing the ELU activation function, the system attained accuracies of up to 99.98% on the N-BaIoT, CICIDS-2017, and ToN-IoT datasets. Its main advantage is extremely high accuracy alongside reduced processing time and improved interpretability for scalable IoT applications [19].

Shaikh et al. (2024) developed an efficient intrusion detection system for DDoS attack identification using a hybrid CNN-LSTM model. In this approach, CNN is employed for spatial feature extraction and LSTM for temporal feature analysis, while an autoencoder is used for dimensionality reduction and SMOTE for data balancing. Evaluated on the real-world CICDDoS2019 dataset, the model achieved 99.86% accuracy. Its main advantage lies in high accuracy combined with simultaneous spatio-temporal feature analysis for DDoS attack detection [20].

Nandanwar et al. (2024) proposed the TL-BiLSTM IoT model for detecting botnet attacks in IoT devices by integrating transfer learning with a CNN-BiLSTM architecture. By leveraging pre-trained model knowledge, this approach achieves high accuracy and reduced training time. Evaluation on the N-BaIoT dataset showed a training accuracy of 99.55% and testing accuracy of 99.52%, outperforming previous methods by 3.2% to 16.07%. The main advantage is its high accuracy and excellent generalizability for scalable IoT network security [21].

Wang et al. (2025) proposed a deep intrusion detection system for IoT networks that integrates CTGAN for synthetic data generation with a deep learning model to enhance attack detection. By simultaneously analyzing the spatial and temporal dimensions of the data, the system effectively manages imbalanced datasets and achieves high accuracy. Evaluation on the UNSW-NB15, CIC-IDS2018, and CIC-IoT2023 datasets demonstrated model accuracies ranging from 99.58% to 99.99%. The main advantage of this approach lies in its ability to handle data imbalance and deliver high multi-class detection accuracy while maintaining efficiency suitable for IoT applications [22].

Logeswari et al. (2025) introduced a two-stage optimized intrusion detection system for IoT, combining intelligent feature selection with classification to achieve high cyberattack detection performance. The system consists of three components: data preprocessing, two-layer feature selection (SDFC), and two-stage classification using LightGBM and XGBoost algorithms. The SDFC algorithm selects only eight effective features through statistical methods and particle swarm optimization. Evaluation on the TON-IoT dataset showed an accuracy of 93.70% with a testing time of 0.1895 seconds, outperforming existing methods. Its main advantage is high accuracy coupled with reduced computational overhead for scalable IoT applications [23].

Raghunath et al. (2025) proposed an intelligent intrusion detection system for IoT that employs Principal Component Analysis (PCA) and Particle Swarm Optimization (PSO) for feature selection and SVM for classification. By reducing the feature set from 41 to 18 and selecting an optimal subset, the system achieved 98.5% accuracy on the NSL-KDD dataset. Its main advantage is reduced computational load while maintaining high accuracy, making it suitable for resource-constrained IoT applications [24].

Shafique et al. (2025) presented a spiking neural network (SNN)-based intrusion detection system for IoT, designed to optimize energy consumption. Comparisons with a CNN model on the NSL-KDD dataset indicated that SNN achieved 75% accuracy and a 72% F1-score, performing close to CNN while consuming up to 4.8 times less energy. The primary advantage of this method is its low energy consumption combined with acceptable accuracy, making it suitable for resource-limited IoT devices [25].

Agarwal et al. (2025) proposed a binary intrusion detection system for IoT communications named BCIDS-IoT, which combines multiple machine learning algorithms. Among them, the ANN model demonstrated the best performance, achieving 95% accuracy on the UNSW-NB15 dataset. By reducing the false positive rate while maintaining high detection rates, the system performs well in large-scale networks, making it a reliable option for IoT security [26].

Das et al. (2025) proposed a lightweight IoT intrusion detection system (IDS) that combines downsized LIDSuFNN and LIDSuCNN models with pruning and quantization techniques. Additionally, CTGAN was employed to generate synthetic data and address data imbalance. Evaluation on several benchmark datasets demonstrated that the models achieved comparable accuracy to the baseline while reducing training time by up to 94% and memory consumption by up to 90%. The main advantage of this approach is the balance between high accuracy and computational efficiency, making it suitable for resource-constrained IoT devices [27].

Paul et al. (2025) introduced a hybrid intrusion detection system for DoS attacks in IoT networks, consisting of a two-stage anomaly detection and signature-based classification process. By utilizing a hybrid model and intelligent feature selection, realistic data were generated in a NetSim simulation environment. Results showed that the model could detect up to 95% of malicious samples (CDS) and classify up to 96% of them (CCS). In addition to high accuracy, the low testing time makes this approach lightweight and scalable for IoT security [28].

Rezaei et al. (2025) presented a federated learning and RNN-based intrusion detection system for IoT that is resilient against malicious attacks such as MPLFA. They introduced the SNCOC defense mechanism, which improves model accuracy by up to 47% by removing compromised nodes. Under normal conditions, the system achieves nearly 99% accuracy and outperforms standard methods such as FedAvg and MKrum under attack scenarios. The main advantage of this approach is its high accuracy and robustness against attacks in distributed IoT environments [29]. Table 1 presents a review and comparison of previous methods.

**Table 1:** A Systematic Review of the Literature on Recent Network-Based Intrusion Detection System Approaches

| Ref | Purpose | Techniques used | Dataset used | Type | Performance evaluation parameter | Result | Limitation |
|---|---|---|---|---|---|---|---|
| [15] | Accurate and efficient intrusion detection for cloud environments | Smart feature selection, weighted hybrid classifier (LSTM, SVM, XGBoost, FLN), CSA optimization | NSL-KDD, Kyoto, CSE-CIC-IDS-2018 | Multiclass classification | Accuracy, Precision, F1-score | Acc-99.99% | No detailed scalability or real-time performance analysis |
| [16] | Lightweight IDS for IoT | DNN + BiLSTM, IPCA for dimensionality reduction, Optuna optimization, dynamic quantization | CIC IDS2017, N-BaIoT, CICIoT2023 | Multiclass classification | Accuracy | Acc- 99.49% | Evaluation limited to small set of datasets |
| [17] | Lightweight ConvNeXt-Sf model for IoT in cloud-edge framework | ConvNeXt + ShuffleNet V2 | BoT-IoT, TON-IoT | Multiclass classification | Accuracy | Accuracy improved up to 6.18% | Limited attack variety in evaluation |
| [18] | DNDF-IDS for IoT anomaly detection | Improved Decision Forest + Neural Networks, feature selection | NSL-KDD, CICIDS2017, UNSW-NB15 | Multiclass classification | Accuracy | Acc- 94.09% | No scalability or real-world deployment analysis |
| [19] | Detect IoT botnet attacks using sequential architecture | Feature selection, OOA-optimized Bi-LSTM, ELU activation | N-BaIoT, CICIDS-2017, ToN-IoT | Multiclass classification | Accuracy | Acc- 99.98% | No detailed scalability or robustness analysis |
| [20] | Detect DDoS attacks with spatial-temporal analysis | CNN for spatial features, LSTM for temporal features, autoencoder, SMOTE | CICDDoS2019 | Multiclass classification | Accuracy | Acc-99.86% | Focused only on DDoS attacks |
| [21] | Detect IoT botnet attacks with transfer learning and CNN-BiLSTM | Transfer learning, CNN-BiLSTM | N-BaIoT | Multiclass classification | Accuracy | Accuracy-99.52% | Evaluation on single dataset |
| [22] | Deep IDS with CTGAN for IoT | CTGAN for synthetic data generation, deep learning | UNSW-NB15, CIC-IDS2018, CIC-IoT2023 | Multiclass classification | Accuracy | Acc-99.58% | Synthetic data quality impact not deeply analyzed |
| [23] | Two-stage optimized IDS for IoT | SDFC two-layer feature selection, LightGBM, XGBoost | TON-IoT | Multiclass classification | Accuracy, Test time | Acc- 93.70%, | Lower accuracy compared to some deep learning methods |

| | | | | | | | | |
|---|---|---|---|---|---|---|---|---|
| [24] | PCA + PSO optimized SVM for IoT IDS | PCA, PSO, SVM | NSL-KDD | Multiclass classification | Accuracy | Acc-98.5% | Single dataset evaluation, limited attack diversity |
| [25] | Energy-efficient IDS for IoT | Spiking Neural Networks (SNN) vs CNN | NSL-KDD | Multiclass classification | Accuracy, F1-score, Energy consumption | Acc-75%, | Lower accuracy than CNN |
| [26] | BCIDS-IoT for binary intrusion detection | Hybrid ML algorithms, ANN | UNSW-NB15 | Binary classification | Accuracy | Acc-95% | Binary detection only, no multi-class support |
| [27] | Lightweight IDS for IoT | LIDSuFNN, LIDSuCNN, pruning, quantization, CTGAN | Multiple public IDS datasets | Multiclass classification | Accuracy, Training time, Memory usage | Near-base accuracy, 94% less training time, 90% less memory usage | Possible accuracy drop in highly complex datasets |
| [28] | Hybrid IDS for DoS detection in IoT | Hybrid anomaly + signature detection, feature selection | NetSim simulated data | Multiclass classification | CDS, CCS | CDS: 95%, CCS: 96% | Simulation-based evaluation only |
| [29] | Federated RNN IDS for IoT | Federated learning, RNN, SNCOC defense | Not specified clearly (IoT datasets) | Multiclass classification | Accuracy | Near 99% in normal, +47% accuracy improvement | Dataset details missing, generalizability not analyzed |

## 3. Fundamental Concepts

In this section, Convolutional Neural Networks (CNNs) and the ConvNeXt-Tiny model, which are employed in the proposed method, are briefly described.

### 3.1. Convolutional Neural Networks (CNNs)

Convolutional Neural Networks (CNNs) are effective and lightweight tools for automatic feature extraction and data classification, with widespread applications across various domains, particularly in cybersecurity [31]. The CNN architecture consists of multiple layers, including input, convolutional, pooling, fully connected, and output layers. Through convolution operations—a mathematical process between matrices—CNNs can extract important and complex features from input data. These features enable the identification of critical patterns in data generated by Internet of Things (IoT) networks [32]. The ability of CNNs to automatically learn features makes them a suitable alternative to traditional feature selection methods. In lightweight intrusion detection systems specifically designed for IoT networks, CNNs contribute to effective feature extraction and classification while reducing computational complexity and resource consumption. This advantage makes CNNs a fundamental component in designing cost-effective and optimized intrusion detection systems for resource-constrained IoT environments [33].

### 3.2. ConvNeXt-Tiny Model

ConvNeXt-Tiny is a lightweight and optimized model from the family of convolutional neural networks, inspired by transformer architectures and designed using advanced techniques. The model is developed to maintain high accuracy while reducing computational complexity and resource consumption, making it highly suitable for resource-constrained applications such as Internet of Things (IoT) networks. ConvNeXt-Tiny has a structure similar to traditional convolutional networks but incorporates modifications in layers and processing mechanisms to provide improved performance in extracting complex spatial features. The model utilizes convolutional blocks with specific configurations, enabling more precise and deeper feature extraction at a lower computational cost. A key feature of ConvNeXt-Tiny is its ability to combine high performance with low resource usage, making it an ideal choice for developing lightweight intrusion detection systems in IoT networks. By maintaining accuracy and leveraging a simpler, optimized structure, the model can be implemented on hardware-constrained devices, enhancing processing speed and reducing energy consumption.

## 4. Proposed Method: CNN + ConvNeXt-Tiny

With the rapid proliferation of Internet of Things (IoT) devices, securing these networks has become critical. Traditional intrusion detection methods often fail to meet the security requirements of IoT networks due to computational limitations and the complexity of dynamic network structures. Therefore, developing lightweight and efficient intrusion detection systems that can operate within the limited resources of IoT devices is essential. Many researchers have proposed deep learning-based approaches to improve attack detection accuracy; however, complex models typically consume significant resources and are not suitable for IoT environments. Additionally, challenges such as processing high-dimensional raw data and the wide variety of botnet attacks have prevented the complete resolution of intrusion detection issues. The objective of this study is to design and implement a lightweight intrusion detection system capable of accurately and rapidly identifying botnet attacks in IoT networks. By leveraging both spatial and temporal features of the data, the system is optimized to maintain high performance while addressing the resource constraints of IoT devices. This paper focuses on developing a lightweight intrusion detection system for IoT networks using a hybrid CNN and ConvNeXt-Tiny model. The main goal is to create an efficient and compact method that can accurately detect intrusion attacks in IoT environments while considering the hardware limitations of IoT devices. The proposed method combines the spatial feature extraction capabilities of Convolutional Neural Networks (CNNs) with the high efficiency of the ConvNeXt-Tiny model, enabling the detection of complex intrusion patterns in network data. This method delivers high accuracy while remaining lightweight and is designed to enhance IoT network security and reduce computational and energy costs.

In this section, the proposed method is presented step by step. The proposed algorithm is explained, and the overall flowchart of the intrusion detection process using the hybrid CNN and ConvNeXt-Tiny model is illustrated. The general stages of the proposed method are shown in Figure 1.

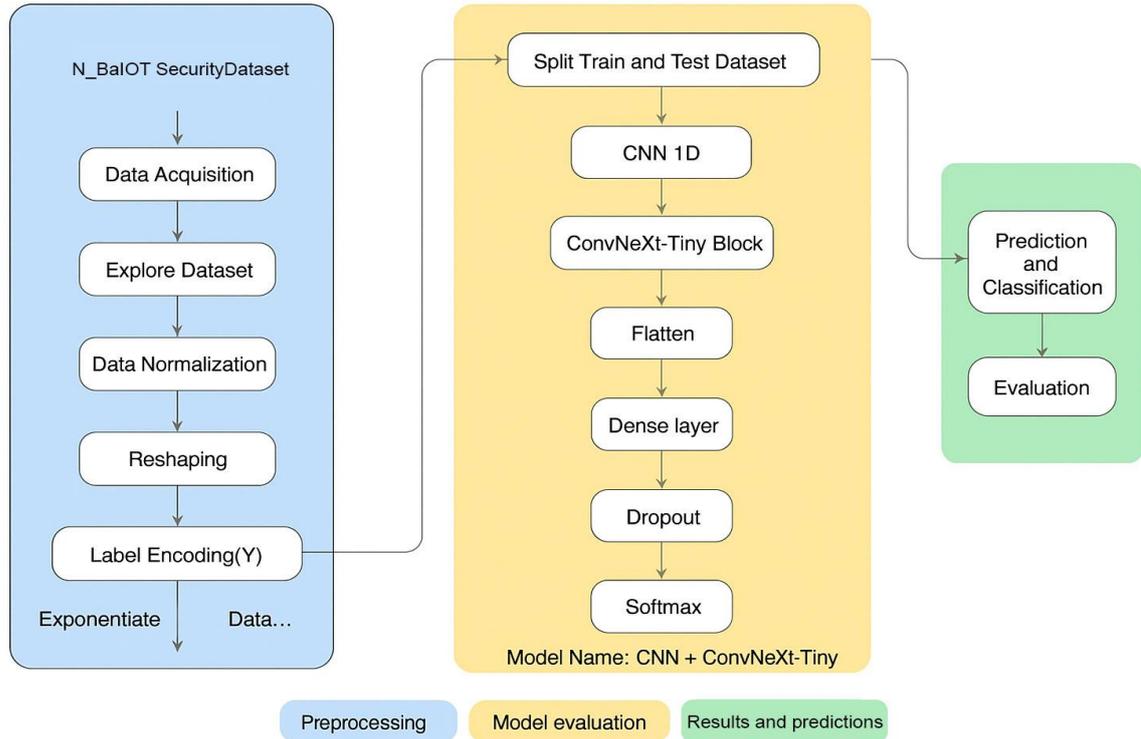

**Figure. 1:** Flowchart of the Proposed Method

### 4.1. Data Preprocessing

In the proposed method based on the hybrid CNN and ConvNeXt-Tiny model for IoT networks, the data preparation process plays a crucial role. This stage involves identifying anomalies and attacks present in the data, transforming the data into a format suitable for deep learning models, and standardizing it to maintain accuracy and comparability. Proper execution of these steps ensures that the input data to the model is appropriate and of high quality, resulting in reliable and valid outputs from the intrusion detection system. In summary, precise data preprocessing forms the foundation for extracting effective and trustworthy results in this type of security system. In this section, the data preprocessing techniques employed in the proposed method are explained in detail.

### 4.1.1. Data Collection

In the proposed method, it is assumed that the collection of network traffic generated by IoT nodes is performed on the client side. A common method for capturing such data is Port Mirroring, where a network switch is configured to send a copy of the incoming and outgoing traffic to a monitoring port, allowing recording and analysis. However, since this paper is based on the analysis of pre-collected and ready-to-use datasets, there is no need to implement direct data collection, and the primary focus is placed on data processing and the design of the intrusion detection framework.

### 4.1.2. Feature Encoding

In this paper, the utilized data consist of a set of numerical features extracted from network packets of IoT devices. These features have already been preprocessed numerically, and no categorical encoding is required. Unlike some methods that employ techniques such as One-Hot Encoding to convert categorical data into binary vectors, in this paper, all features are defined as continuous or discrete numerical values and can therefore be used directly for training the proposed method. Since the dataset contains no textual or categorical features requiring encoding, there was no need to apply methods such as Label Encoding or use functions like get_dummies. This choice reduces processing complexity and increases model training speed. Accordingly, the feature encoding stage in this study primarily involves standardizing numerical features through normalization using StandardScaler, which is described in the following section.

### 4.1.3. Data Standardization

To improve data quality and unify the scale of numerical features extracted from IoT network packets, standardization was applied in this paper. This statistical method transforms the original values of each feature to a new range with a mean of zero and a standard deviation of one, allowing direct and fair comparison of features with different scales. Standardization is performed by subtracting the mean of each feature from its data value and dividing the result by the standard deviation of the same feature, as expressed in Equation (1):

$$Z = \frac{(x - \mu)}{\sigma} \qquad (1)$$

Where μ represents the mean of the feature distribution, σ is the standard deviation of the feature, and Z denotes the standardized score.

The Z value indicates the number of standard deviations a sample is away from the mean. For instance, *Z=3* indicates that the sample is three standard deviations above the mean. Through this process, the data distribution is transformed towards a normal distribution with a mean of zero and a standard deviation of one, which significantly contributes to improving the performance of machine learning models.

### 4.2. Dataset Analysis

In the context of IoT network security, datasets often contain precise and detailed records that cover a diverse range of features related to both normal and malicious device behaviors. Analyzing these data provides the foundation for designing and training effective models for anomaly and malicious activity detection. Therefore, a thorough understanding of the data structure and attack patterns plays a crucial role in enhancing the performance of intrusion detection systems. In this paper, a dataset derived from botnet attacks in IoT environments is utilized, initially introduced by Mirsky et al. [30], and has been widely employed in numerous studies as a reliable source for evaluating intrusion detection systems. This dataset, containing real network traffic data with both normal and malicious activities, allows for a more accurate assessment of the performance of our proposed method. The dataset includes various types of attacks, which are detailed in Table 2. These attacks were used for training a multi-class classification method employing transfer learning. This information was taken into account during the analysis to enable the system to accurately identify and differentiate each type of attack, particularly within IoT network environments.

**Table 2:** Systematic presentation of the number of "Detection of IoT botnet attacks N_BaIoT" security dataset

| Device Name | Benign Samples (Train) | Benign Samples (Test) | Attack Samples (Test) |
|---|---|---|---|
| Provision Security Camera | 43,507 | 18,647 | 981 |
| Philips Baby Monitor | 122,668 | 52,572 | 2,767 |
| Ecobee Thermostat | 9,179 | 3,934 | 207 |
| Danmini Doorbell | 34,683 | 14,865 | 782 |
| Samsung Webcam | 36,505 | 15,645 | 823 |

## 4.3. Architecture and Formulation of the Proposed Method

In this paper, a lightweight and effective model for intrusion detection in Internet of Things (IoT) networks is introduced, leveraging a combination of Convolutional Neural Networks (CNNs) and the modern ConvNeXt-Tiny architecture. The proposed method is designed to accurately and efficiently detect attacks in network traffic data, with the capability of automatic feature extraction and complex pattern recognition, while remaining computationally suitable for resource-constrained environments such as fog and edge nodes. Figure 2 presents a simplified view of the overall structure of the proposed method as depicted in Figure 4.

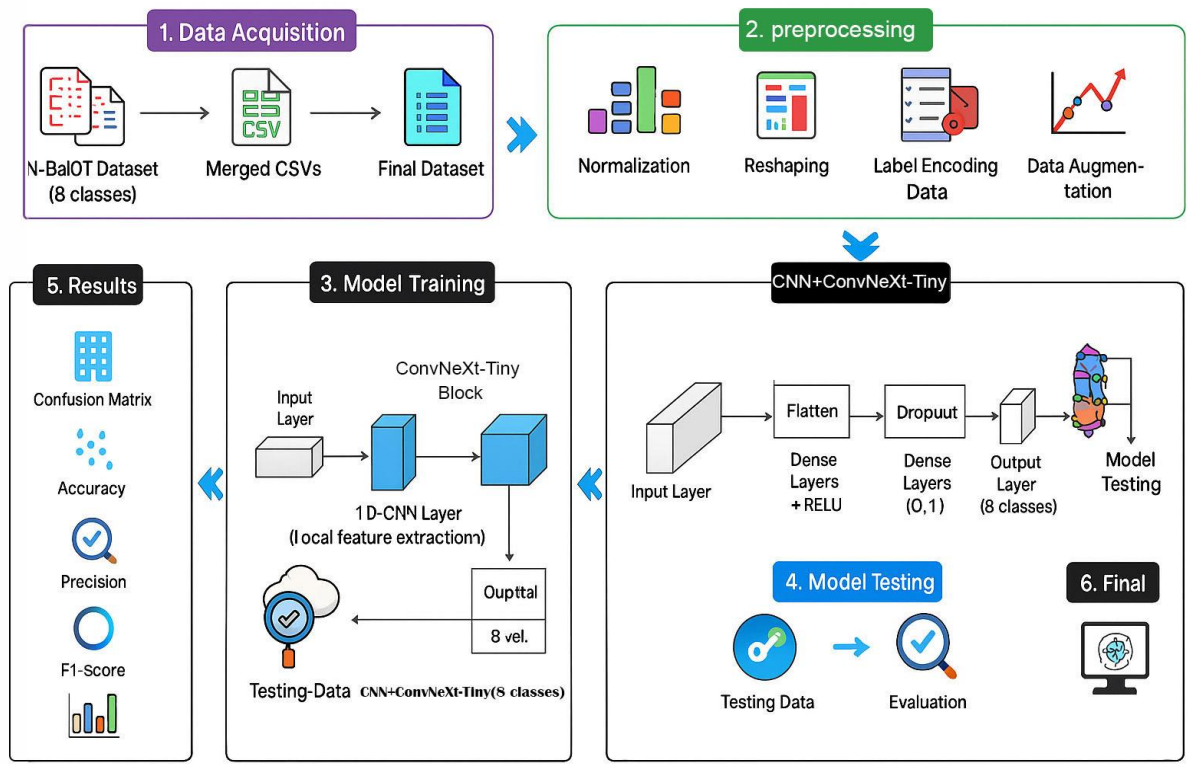

**Figure. 3** Simplified View of the Proposed Hybrid CNN + ConvNeXt-Tiny model

The architecture of the proposed method, along with the role of each layer, is described step by step.

Input Data → Conv1D Layer → ConvNeXt-Tiny Block → Flatten → Dense Layer → Dropout → Output Layer (Softmax)

First Layer (Data Preprocessing): Raw network data are initially prepared for model input using normalization and appropriate reshaping techniques.

Second Layer (1D Convolutional Layer, 1D-CNN): This layer is responsible for extracting initial local features from network traffic. Using small-sized filters, information related to the basic structure of data packets is identified.

Third Layer (ConvNeXt-Tiny Block): Following the initial CNN layer, the data enter a compressed and optimized version of the ConvNeXt-Tiny architecture. This block leverages transformer-like structures and optimizations for both visual and sequential processing to extract high-level features. Since ConvNeXt-Tiny is designed for high efficiency on resource-constrained hardware, it is an ideal choice for IoT applications.

Fourth Layer (Flatten): The output of the ConvNeXt-Tiny block is converted into a one-dimensional vector to be suitable for the fully connected layers.

Fifth Layer (Dense / Fully Connected Layers): This stage includes one or more fully connected layers with ReLU activation functions to aggregate and process features and learn complex non-linear relationships among them.

Sixth Layer (Dropout Layer): To prevent overfitting, a dropout layer with a rate of 0.1 is applied, which deactivates a portion of neurons during each training step.

Final Layer (Softmax Output Layer): Finally, the output layer with a Softmax activation function determines the probability of the sample belonging to each attack class or normal traffic.

The mathematical formulation of the proposed method is as follows:

Let x represent the input data. The model input is defined as a three-dimensional tensor, as expressed in Equation (2):

$$x \in \backslash R^{(n \times m \times 1)} \tag{2}$$

Where *n* represents the number of time steps and *m* denotes the number of features.

The one-dimensional convolution operation in the second layer can be expressed as follows: To extract initial local features from traffic data, a one-dimensional convolutional layer with *F* filters and a kernel size of *K* is applied. The operation of this layer is defined by Equation (3):

$$Y_{i,j} = \sum_{k=1}^{K} w_k \cdot x_{i,j+k-\left[\frac{K}{2}\right]} \tag{3}$$

Where $w_k$ denotes the weights of the convolutional kernel. The final output, with the ReLU activation function, is defined as in Equation (4):

$$Z_{i,j} = \sigma(Y_{i,j} + b), \quad Where\ \sigma(x) = \max(0, x) \tag{4}$$

The third layer of the proposed method consists of the ConvNeXt-Tiny block, where the output of the CNN layer is fed into the ConvNeXt-Tiny block. This block includes multiple stages of modern ResNet-like blocks and utilizes normalization layers and depthwise convolutions. In summary, high-level features with spatio-temporal distribution are extracted by ConvNeXt-Tiny, as defined in Equation (5):

$$Z_{ConvNeXt} = ConvNeXtTiny(Z) \tag{5}$$

In the fourth layer, the Flatten and Dense layers, the output of the ConvNeXt-Tiny block is transformed into a one-dimensional vector $h_1$ using Flatten, as expressed in Equation (6):

$$h_1 = Flatten(Z_{ConvNeXt}) \tag{6}$$

Then, in the fifth layer, Fully Connected layers are used to learn complex patterns, as expressed in Equations (7) and (8):

$$z_1 = W_1 h_1 + b_1, \quad a_1 = ReLU(z_1) \tag{7}$$

$$z_2 = W_2 a_1 + b_2, \quad a_2 = ReLU(z_2) \tag{8}$$

The sixth layer includes a Dropout layer. To prevent overfitting, a Dropout layer with a rate of p=0.1 is applied, as expressed in Equation (9):

$$d = Dropout(a_2, p = 0.1) \tag{9}$$

Finally, a Softmax layer is used to produce a probability distribution over the output classes, as expressed in Equations (10) and (11):

$$z_3 = W_3 d + b_3 \tag{10}$$

$$\hat{y} = Softmax(z_3) = \frac{e^{z_3}}{\sum_i e^{z_{3i}}} \tag{11}$$

Where $\hat{y}$ is the final predicted vector for attack and normal traffic classes.

In summary, the proposed method in this study combines convolutional layers (CNN) with the lightweight ConvNeXt-Tiny architecture to automatically extract effective features from

---

Algorithm: Proposed Lightweight CNN + ConvNeXt-Tiny method

Input: raw network traffic data, preprocessing pipeline, feature set, training data
Output: classification of normal traffic and different IoT attack types

Function: Build CNN + ConvNeXt-Tiny model (train_data, labels)

1. Preprocessing Layer: normalize input features and reshape raw traffic into model-compatible format
2. Create a sequential model
3. Conv1D Layer: filters = 64, kernel size = 5, padding = same, activation = ReLU, input shape = preprocessed features
4. ConvNeXt-Tiny Block: optimized lightweight ConvNeXt-Tiny architecture for extracting high-level features
5. Flatten Layer: convert ConvNeXt-Tiny output into a 1-D vector
6. Dense Layer: 128 units, activation = ReLU
7. Dense Layer: 64 units, activation = ReLU
8. Dropout Layer: rate = 0.1 to prevent overfitting
9. Output Dense Layer: number of units = number of attack/normal classes, activation = Softmax
10. Compile the model → optimizer = Adam, loss = categorical_crossentropy, metrics = accuracy, AUC
11. Train the model → training set, validate on validation set

---

network traffic data. The CNN layer extracts initial local features, while ConvNeXt-Tiny, with its modern and optimized structure, identifies more complex relationships and high-level features. Dense layers are used to combine features and learn non-linear patterns, and the Dropout layer enhances model generalization by reducing overfitting. Finally, the output layer with the Softmax activation function produces a probabilistic distribution over the classes, enabling accurate multi-class classification of attacks in IoT networks. The following algorithm illustrates the design and training process of the proposed CNN + ConvNeXt-Tiny model for network traffic classification.

## 5. Simulation and Results

The objective of this section is to provide a concise analysis of the experimental settings, describe the dataset, and present the results of the proposed hybrid deep learning method on a benchmark dataset, with a focus on various performance evaluation metrics.

### 5.1. Experimental Settings

The experiments were conducted using an ASUS TUF Gaming F15 (FX506LHB) laptop equipped with a 10th-generation Intel Core i5 processor, 8 GB of RAM, 512 GB of storage, and Windows 11 operating system. The laptop also featured an NVIDIA GTX1650 GDDR6 graphics card with 4 GB of memory, which delivered satisfactory performance throughout the experiments. For data exploration and analysis, various data analysis frameworks were employed, including Pandas, Numpy, Seaborn, Matplotlib, and Scikit-learn. The RAM capacity limitation of 8 GB was also taken into account during the experimental process.

### 5.2. Dataset Description

Data mining in the field of cybersecurity is typically performed on datasets containing numerous records of security-related features and associated details. These datasets can be used to build security models aimed at detecting malicious or anomalous activities. Therefore, understanding the fundamental characteristics of cybersecurity data and identifying attack patterns is essential for effective anomaly detection.

In this paper, the "Detection of IoT Botnet Attacks (N_BaIoT)" dataset was used as the primary data source. This dataset was first introduced by Mirsky and Meidan [30] and was generated by capturing network traffic from five selected devices out of nine different IoT devices (as detailed in Table 2). The N_BaIoT dataset addresses the lack of publicly available datasets for IoT botnet research. It is multivariate and sequential, comprising a total of 115 continuous numerical features. The data are categorized into normal traffic and 10 types of attacks executed by two well-known botnets, Mirai and BASHLITE. In this paper, normal traffic and seven attack types were analyzed (as shown in Table 3).

**Table 3:** Overview of the Dataset, Attack Types, and Total Number of Samples per Target Class

| Mirai and BASHLITE | Type of attack | Description | Number of instances |
|---|---|---|---|
| BASHLITE | Gafgyt combo | Sending spam data to a network—transmitting unsolicited or unwanted messages or advertisements to a network, often with the intention of overwhelming the system or spreading malware | 15,345 |
| BASHLITE | Gafgyt Scan | Network scanning for attacking systems—examininga network to identify vulnerabilities or potential targets for a botnet attack | 14,648 |
| BASHLITE | Gafgyt junk | Sending spam data-distributed or transmitting unsolicited or unwanted messages or advertisements through various means such as email and text messages | 15,449 |
| Mirai | Mirai UDP | Scanning the network for victim devices—the process of searching a network for a specific device that is the intended target of an attack. The scanning is to identify the IP address, open port and vulnerabilities of the victim device | 15,602 |
| Mirai | Mirai plain UDP | UDP flooded by optimizing seeding packets per second—UDP is flooded by sending an excessive number of seeding packets per second in an attempt to optimize the process | 15,304 |
| Mirai | Mirai syn | Sending a flood of synchronization—overwhelming a network with a large number of synchronization messages, often with the intention of disrupting the normal operation of the system or causing a denial of service attack | 16,436 |
| Mirai | Mirai ack | Sending a flood of acknowledgment—overwhelming a network with a large number of acknowledgment messages, often with the intention of disrupting | 15,138 |
| None | Benign | Unharmful network data—it is legitimate data that is not intended to cause harm or damage to a system | 15,538 |

### 5.3. Performance Evaluation Metrics

To assess the quality and accuracy of the intrusion detection system, metrics such as Precision, Recall, F1-Score, and Accuracy were used. Considering that the current problem is a multi-class classification task with discrete model outputs, a metric for comparing these classes is necessary [34, 35]. In this context, we first introduce the concept of the Confusion Matrix.

**1. Confusion Matrix:** A table that compares the true labels of the samples with the labels predicted by the model. This matrix includes values that indicate the correctness of predictions. Specifically:
- True Positive (TP): The number of positive samples correctly predicted by the model.
- True Negative (TN): The number of negative samples correctly identified.
- False Positive (FP): The number of negative samples incorrectly predicted as positive (Type I error).
- False Negative (FN): The number of positive samples incorrectly predicted as negative (Type II error).

**2. Precision:** Also known as Positive Predictive Value (PPV), precision measures the proportion of correctly predicted positive instances according to Equation (12). Precision is defined as the ratio of true positives to all positive predictions made by the model.

$$Precision = \frac{TP}{TP + FP} \tag{12}$$

That is, the number of correctly identified attacks divided by the sum of correctly identified attacks and false positives (non-attacks incorrectly classified as attacks).

**3. Recall:** Also known as the True Positive Rate (TPR), recall represents the percentage of samples correctly classified by the algorithm, as defined in Equation (13). Recall is calculated as the ratio of true positives to the total number of actual positive samples in the dataset.

$$Recall = \frac{TP}{TP + FN} \tag{13}$$

That is, the number of correctly identified attacks divided by the sum of correctly identified attacks and non-attack samples incorrectly classified as attacks.

**4. F1-Score:** The F1-score is a combined metric that balances Precision and Recall, calculated as their harmonic mean according to Equation (14).

$$F1 - score = \frac{2}{\left[\left(\frac{1}{Precision}\right) + \left(\frac{1}{Recall}\right)\right]} \tag{14}$$

**5. Accuracy:** Accuracy is an important metric for evaluating the overall performance of the algorithm, representing the ratio of correctly classified samples to the total number of samples, as defined in Equation (15).

$$Accuracy = \frac{TP}{TP + FN + TN + FP} \tag{15}$$

These performance metrics serve as fundamental tools for evaluating algorithm effectiveness across various domains, particularly in research contexts.

## 5.4. Results

In this section, the proposed method for intrusion detection in the Internet of Things (IoT) is evaluated. The proposed method was employed to describe the evaluation and system testing procedure, aiming to detect botnet attacks using the "Detection of IoT Botnet Attacks (N_BaIoT)" security dataset. The dataset was split such that 20% of the samples were used for testing, while the remaining 80% were allocated for training and validation. The training set was further divided, with a portion used as a validation set to assess the model's performance on unseen data and to optimize hyperparameters. Our proposed method, ConvNeXt-Tiny+ CNN, a multi-class deep neural network, achieved an exceptionally high accuracy of 99.63% on the test data, with results summarized in Table 4. The high accuracy on the training set demonstrates the model's capability to effectively learn from the training data and correctly predict class labels, while the high accuracy on the test set indicates the proposed method's ability to generalize to new and unseen data. These results confirm that the proposed method can provide precise predictions on real-world and previously unknown samples.

**Table 4:** Performance Evaluation of the Proposed Method

| Dataset | Accuracy (%) | Loss | Precision (%) | Recall (%) | AUC (%) |
|---|---|---|---|---|---|
| Train set | 99.67 | 0.0119 | 99.76 | 99.75 | 100.00 |
| Validation set | 99.53 | 0.0693 | 99.54 | 99.53 | 99.90 |
| Test set | 99.63 | 0.0107 | 99.63 | 99.63 | 100.00 |

With the aim of developing a lightweight and well-fitted intrusion detection system for Internet of Things (IoT) networks, we designed a hybrid model based on CNN and ConvNeXt-Tiny. This method achieved training and validation accuracies of 99.63% and 99.53%, respectively, as illustrated in Figure 3(a). In a well-fitted model, the validation accuracy is typically slightly lower than the training accuracy, indicating the model's ability to generalize to new data.

In the proposed method, the CNN component is responsible for extracting local features from the input data, while the ConvNeXt-Tiny module, with its modern and optimized structure, enhances the model's capacity to learn deep and complex features. The outputs of these two components are merged and fed into a fully connected layer for final classification. Model training is performed using backpropagation and gradient descent, with the categorical cross-entropy loss function.

Figure 3(b) shows the reduction of training and validation loss over epochs, both decreasing steadily and stabilizing within a consistent range, with only minor differences between their final values. This indicates that our method avoids both overfitting and underfitting, demonstrating stable learning behavior.

Figure 3(c) presents the Precision of the proposed method for training and validation across epochs. High values in this metric indicate the method's effectiveness in correctly identifying true positive samples while minimizing false positives.

Figure 3(d) illustrates the Recall, reflecting the method's high capability in detecting all positive samples accurately.

Finally, Figure 3(e) displays the learning rate set at a constant value of 0.001. As a critical hyperparameter, the learning rate plays a key role in the speed and stability of model convergence. The consistent rate throughout training allows the proposed method to move steadily toward the loss function's minimum, providing stable and reliable performance.

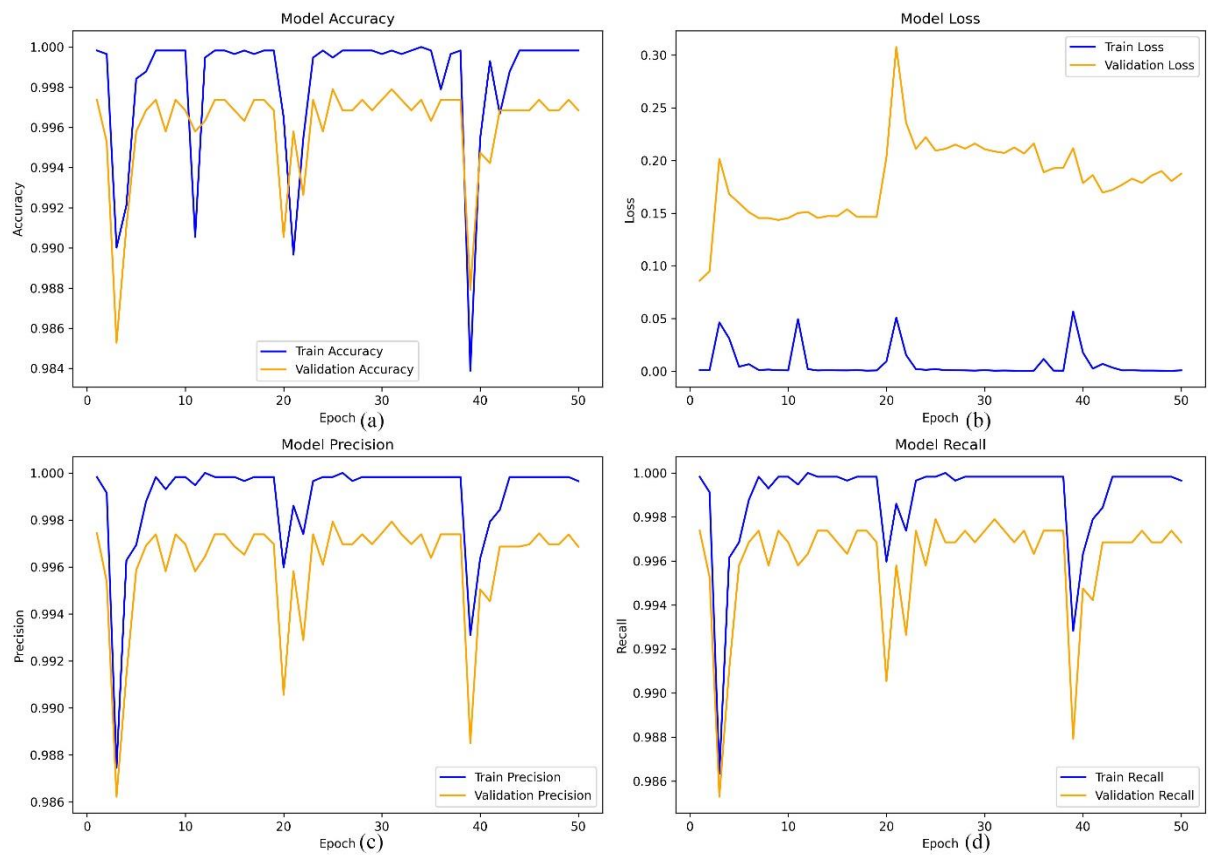

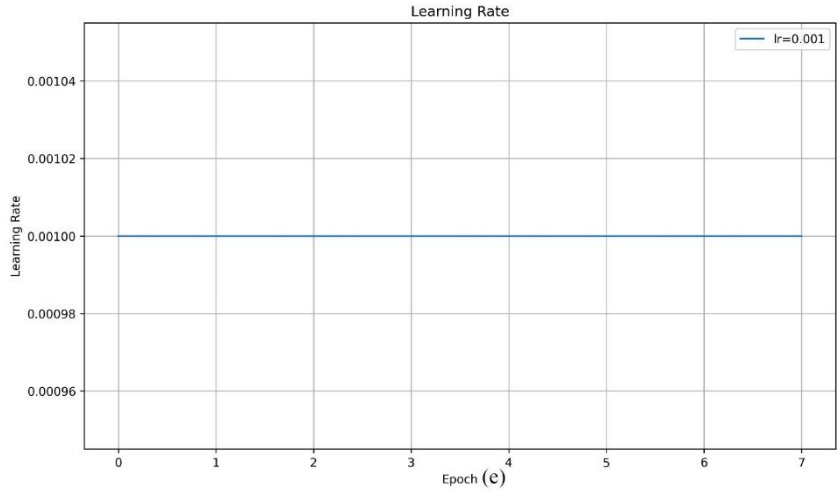

**Figure. 3:** (a) Training and validation accuracy, (b) Training and validation loss, (c) Precision during training and validation, (d) Recall during training and validation, (e) Learning rate.

Table 5 presents the evaluation results of the proposed method with hyperparameter optimization for seven attack types and one normal class in IoT networks. The performance metrics include Precision, Recall, and F1-Score, ranging from 0 to 1. High values of Precision and Recall indicate that the CNN and ConvNeXt-Tiny hybrid model is highly capable of accurately distinguishing between different attack types and normal traffic, while a high F1-Score reflects a proper balance between these two metrics. In Figure 4, the Precision, Recall, and F1-Score values of the proposed method for each of the seven attack types and the normal class are illustrated. The results demonstrate that the proposed method achieved perfect performance (1.00) in all metrics for several classes, including GAFGYT_COMBO, GAFGYT_JUNK, GAFGYT_SCAN, and MIRAI_UDP_PLAIN, while other classes also exhibited near-optimal performance.

**Table 5:** Experimental results of the proposed method on various performance evaluation metrics

| Class | Precision | Recall | F1-Score |
|---|---|---|---|
| BENIGN | 1 | 0.98 | 0.99 |
| MIRAI_UDP | 0.99 | 1 | 0.99 |
| GAFGYT_COMBO | 1 | 1 | 1 |
| GAFGYT_JUNK | 1 | 1 | 1 |
| GAFGYT_SCAN | 1 | 1 | 1 |
| MIRAI_ACK | 0.98 | 1 | 0.99 |
| MIRAI_SYN | 1 | 0.99 | 0.99 |
| MIRAI_UDP_PLAIN | 1 | 1 | 1 |

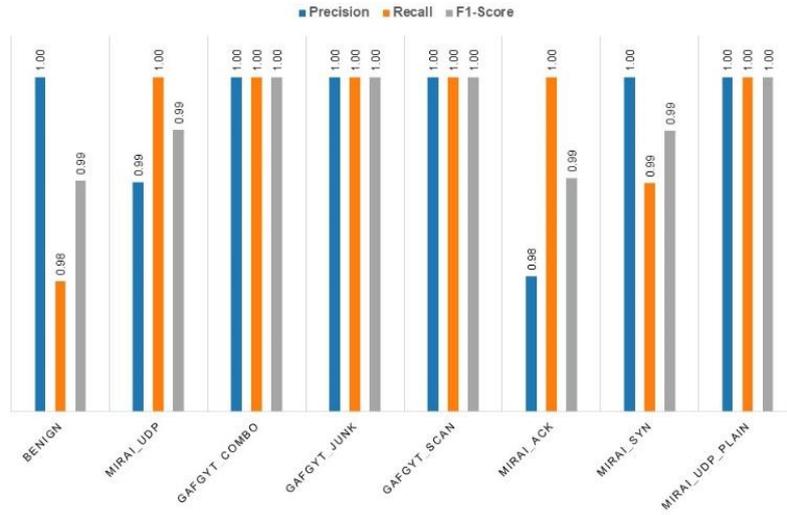

**Figure. 4:** Performance evaluation metrics of the proposed method, including Precision, Recall, and F1-Score.

We also calculated the micro, macro, and weighted averages to evaluate the overall performance of the proposed method, as shown in Table 6. The micro average indicates that each sample or prediction is equally weighted. In contrast, the macro average assesses the overall performance of the classifier without considering data imbalance, while the weighted average takes the effect of class imbalance into account. The macro average score of 99.39% indicates that the proposed method performs very well across all classes, and its accuracy in smaller classes is not affected by larger classes.

The classification performance of the combined CNN and ConvNeXt-Tiny model was also evaluated using confusion matrices, shown in Figure 6. These matrices are used in multi-class classification to compare predicted labels with true labels. Figure 5a shows the confusion matrix for the training set, indicating the number of correct and incorrect predictions for each class. The high number of correct predictions demonstrates that the proposed method has accurately learned the patterns in the training data. Figure 5b shows the confusion matrix for the validation set, which is used to evaluate and compare the model's performance on unseen data and to optimize hyperparameters. Figure 5c presents the confusion matrix for the test set, assessing the proposed method's ability to correctly identify new and unseen samples. The results of these three matrices indicate that the proposed method has a high capability to correctly detect attacks and normal traffic across the training, validation, and test stages in IoT networks.

**Table 6:** Classification results for detecting botnet attacks on the test dataset

| Macro precision (%) | Macro recall (%) | Macro F1-score (%) | Macro average (%) |
|---|---|---|---|
| 99.39 | 99.39 | 99.39 | 100 |
| Macro precision (%) | Macro recall (%) | Macro F1-score (%) | Micro average (%) |
| 99.39 | 99.39 | 99.39 | 99.89 |
| Macro precision (%) | Macro recall (%) | Macro F1-score (%) | Weighted average (%) |
| 99.39 | 99.39 | 99.39 | 99.89 |

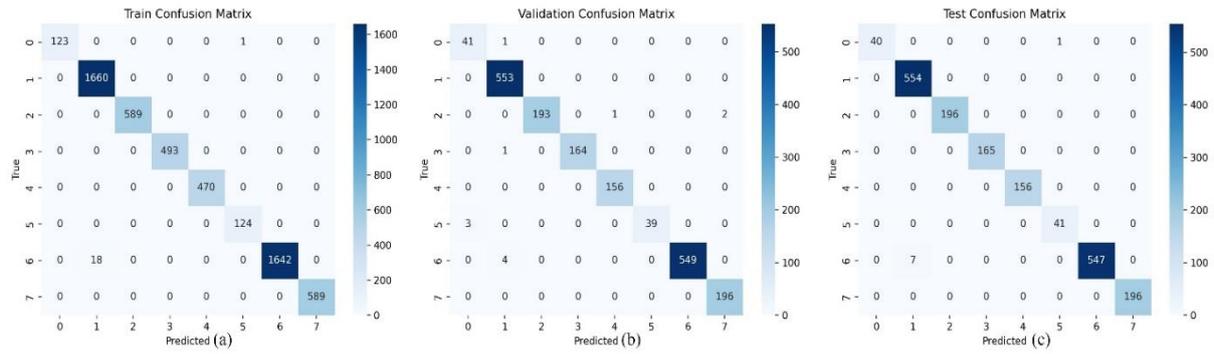

**Figure. 5:** Confusion matrices of the proposed method for different target classes: a) training set, b) validation set, c) test set

The Receiver Operating Characteristic (ROC) curve shown in Figure 6 graphically illustrates the relationship between the True Positive Rate (TPR) and False Positive Rate (FPR) for the eight classes in the IoT intrusion detection task. As observed, the Area Under the Curve (AUC) for all classes is 1.00, indicating the flawless performance of our proposed method in distinguishing between attack and normal traffic classes. The ROC curve is a key tool for evaluating the ability of the proposed method to differentiate between true positives and false positives. Examining this curve allows the determination of an optimal decision threshold that balances sensitivity and specificity. The results demonstrate that our lightweight proposed method, optimized for deployment in resource-constrained IoT environments, achieves very high accuracy and reliability in detecting both attacks and normal traffic. For a more comprehensive performance analysis, in addition to common metrics like Accuracy and Precision, a set of supplementary indicators was employed, including the Matthews Correlation Coefficient (MCC), True Negative Rate (TNR), Negative Predictive Value (NPV), False Positive Rate (FPR), False Discovery Rate (FDR), False Omission Rate (FOR), and False Negative Rate (FNR).

In multiclass classification, the MCC value for each class is calculated individually as a binary classification problem (one class versus all others), and then the average is considered as the overall metric. The MCC ranges from 1 (perfect performance) to -1 (completely incorrect classification), and the values obtained in this study are close to 1, indicating the robustness and reliability of the proposed method across diverse data types.

The Matthews Correlation Coefficient (MCC) is defined by Equation (16):

$$MCC = \frac{(TP \times TN) - (FP \times FN)}{\sqrt{(TP + FP)(TP + FN)(TN + FP)(TN + FN)}} \tag{16}$$

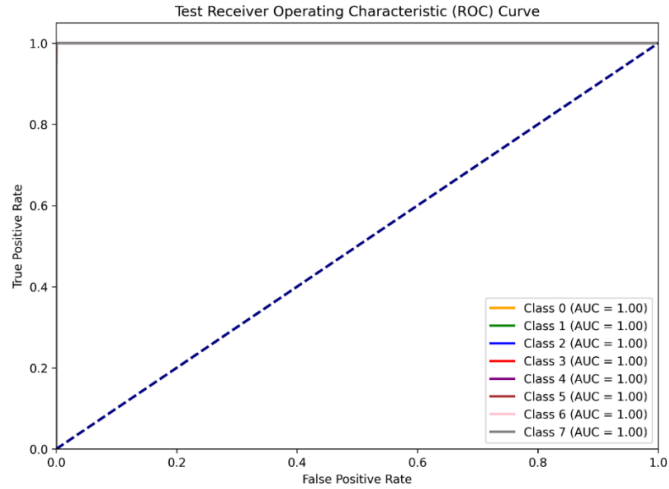

**Figure. 6:** ROC curve for the proposed method: based on the calculated True Positive Rate (TPR) and False Positive Rate (FPR) values.

Our proposed lightweight hybrid method, leveraging CNN and ConvNeXt-Tiny for intrusion detection in IoT networks, has achieved highly stable and reliable performance. This approach attained an impressive Matthews Correlation Coefficient (MCC) of 0.9993 across all target classes. To assess the robustness of the proposed method and compare it with other approaches, a set of critical performance metrics—including False Negative Rate (FNR), False Omission Rate (FOR), False Discovery Rate (FDR), False Positive Rate (FPR), Negative Predictive Value (NPV), and True Negative Rate (TNR)—were computed, with results presented in Table 7. As observed, the proposed method exhibits very low error rates and high accuracy across all classes, particularly in detecting benign samples (BENIGN) and common attacks such as MIRAI and GAFGYT. This highlights the method's capability to precisely identify attacks while minimizing errors, which is crucial for maintaining the security of IoT networks. These results confirm the high stability and reliability of the proposed approach in complex and heterogeneous IoT environments.

**Table 7:** Stability Analysis of the Proposed Method

| Class | TNR | NPV | FPR | FDR | FOR | FNR |
|---|---|---|---|---|---|---|
| BENIGN | 0.999463 | 0.999463 | 0.000537 | 0.02439 | 0.000537 | 0.02439 |
| MIRAI_UDP | 1.000000 | 1.000000 | 0.000000 | 0.00000 | 0.000000 | 0.00000 |
| GAFGYT_COMBO | 1.000000 | 1.000000 | 0.000000 | 0.00000 | 0.000000 | 0.00000 |
| GAFGYT_JUNK | 1.000000 | 1.000000 | 0.000000 | 0.00000 | 0.000000 | 0.00000 |
| GAFGYT_SCAN | 1.000000 | 1.000000 | 0.000000 | 0.00000 | 0.000000 | 0.00000 |
| MIRAI_ACK | 0.999463 | 0.999463 | 0.000537 | 0.02439 | 0.000537 | 0.02439 |
| MIRAI_SYN | 1.000000 | 1.000000 | 0.000000 | 0.00000 | 0.000000 | 0.00000 |
| MIRAI_UDP_PLAIN | 1.000000 | 1.000000 | 0.000000 | 0.00000 | 0.000000 | 0.00000 |

In addition to evaluating the overall performance of the proposed method, the metrics used in the analysis provide insights into specific classes that face particular challenges in classification. This information is highly useful for optimizing the hyper parameters of the proposed method.

The True Negative Rate (TNR) for each class, shown in Figure 7, reflects the model's ability to correctly identify negative samples. Higher values indicate better performance of the proposed method in accurately detecting non-malicious samples. TNR values close to zero or below in some classes suggest minor challenges in identifying negative instances, though performance remains very strong in most classes.

Figure 8 illustrates the Negative Predictive Value (NPV) for each class, representing the method's ability to correctly predict negative samples. Here too, the proposed approach demonstrates satisfactory performance across most classes, with minor fluctuations in some cases that can be further improved.

Figure 9 presents the False Positive Rate (FPR), indicating errors where benign communications are mistakenly identified as attacks. Our method achieves very low FPR in most classes, reflecting a reduction in false alarms and enhanced accuracy of the intrusion detection system.

The False Discovery Rate (FDR), shown in Figure 10, demonstrates that the method accurately identifies positive samples, with values near zero for most classes, indicating excellent performance.

Finally, Figure 11 shows the False Omission Rate (FOR), reflecting the method's ability when correctly identifying negative samples is more critical than positive ones. This metric also confirms the high capability of the proposed method in detecting negative instances and reducing omission errors.

These metrics help pinpoint specific aspects of the proposed method's performance that require improvement and guide the adjustment and optimization of hyperparameters to achieve more stable and accurate operation in diverse and complex IoT network environments.

Overall, our lightweight hybrid method successfully provides high performance in detecting and classifying various IoT botnet attacks. The method achieves high precision, recall, and F1 scores across all attack classes, maintaining a remarkable balance between overall accuracy and the ability to identify true positive instances. Confusion matrices and complementary metrics demonstrate that the proposed method effectively reduces false positives and omission errors, contributing to the optimization and enhancement of the system's capabilities.

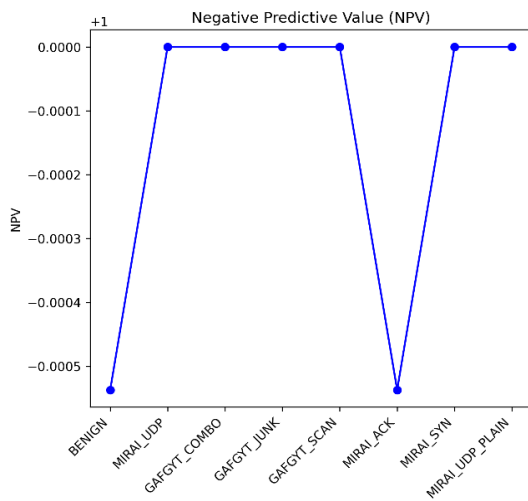 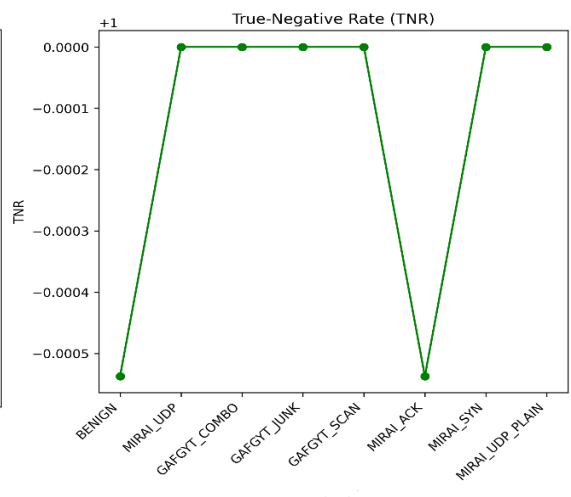

**Figure. 7:** True Negative Rate (TNR) for each class

**Figure. 8:** Negative Predictive Value (NPV) for each class

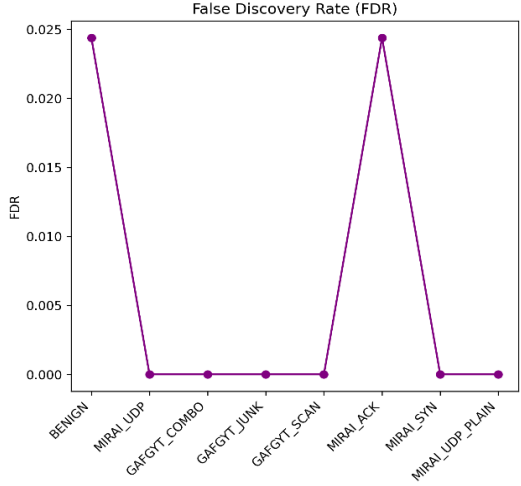
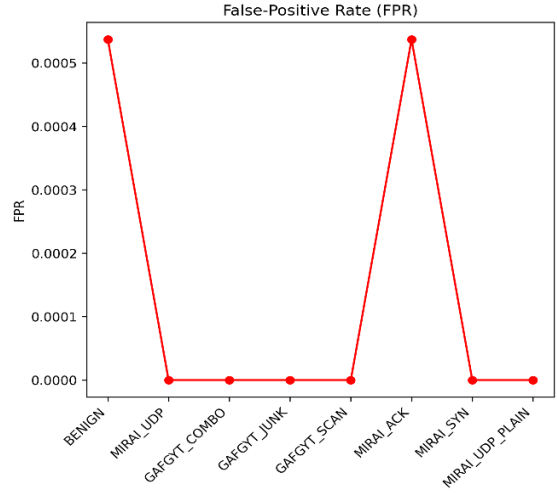

**Figure. 9:** False Positive Rate (FPR) for each class     **Figure. 10:** False Discovery Rate (FDR) for each class

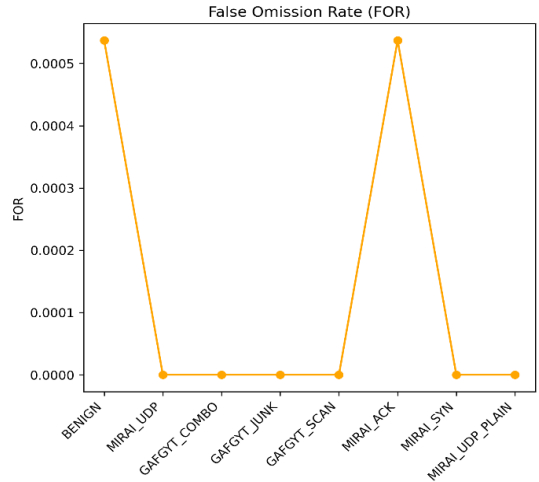

**Figure. 11:** False Omission Rate (FOR) for each class

## 5.5. Complexity of the Proposed Method
In this section, the time and space complexity of the proposed method are calculated and analyzed.

### 5.5.1. Time Complexity
The time complexity of the proposed method mainly depends on the number of trainable parameters, which are influenced by hyperparameters such as the number of filters in the convolutional layers, kernel sizes, the number of units in ConvNeXt-Tiny, and the units in the dense layers. This complexity can be estimated using a formula that incorporates these hyperparameters, whose parameters include:

$$\text{num\_params} = (\text{num\_filters} * \text{kernel\_size} + 1) * \text{num\_features} + \text{num\_features}^2 + \text{num\_features} * \text{num\_timesteps} * (\text{num\_timesteps} - 1) * \text{convnext\_units} + \text{num\_features} * \text{dense\_units} + \text{dense\_units} + \text{num\_classes}$$

- Number of filters (num_filters): The number of filters in the convolutional layers
- Kernel size (kernel_size): The size of the convolutional kernels
- Number of features (num_features): The number of input features

- Number of time steps (num_timesteps): The number of time steps in the input sequence
- ConvNeXt-Tiny units (convnext_units): The number of main units in the ConvNeXt-Tiny structure
- Dense units (dense_units): The number of units in the dense layers
- Number of classes (num_classes): The number of output classes

In practice, the time complexity is approximately proportional to the number of trainable parameters. The most computationally intensive part during training is the gradient computation, which is performed at each training step and requires evaluating the model at each stage. Additionally, the number of training steps depends on the number of epochs and the dataset size, which also affects the time complexity.

Reducing the number of parameters in the proposed method through the lightweight ConvNeXt-Tiny structure and applied optimizations has significantly decreased the training and inference time. This feature enables the system to operate efficiently with fast response times in resource-constrained and distributed IoT network environments. A summary of this analysis is presented in Table 8.

**Table 8:** Computational Time Analysis: Training, Validation, and Testing Performance on the Dataset

| Dataset | Computational time |
| --- | --- |
| Train set | 56.21 s (6.30 ms/step) |
| Validation set | 0.33 s (5.52 ms/step) |
| Test set | 0.33 s (5.52 ms/step) |

### 5.5.2. Space complexity

The Space complexity of the proposed intrusion detection system is directly related to the number of trainable parameters in the model. Each parameter requires memory for storage. Additionally, memory is consumed for processing the input data from IoT networks, intermediate activations, and gradients during the training phase. Memory usage increases with the size of the input data, the complexity of the hybrid model architecture, and the number of training steps. Considering the resource constraints in IoT environments, efficient management of spatial complexity is highly important. However, the use of optimized algorithms and specialized hardware, such as low-power and lightweight GPUs, can help mitigate these limitations. Moreover, techniques like weight regularization and early stopping contribute to reducing the number of parameters and preventing overfitting, thereby optimizing the spatial complexity of the proposed method.

### 5.6. Comparison with Previous Methods

In this section, the proposed CNN + ConvNeXt-Tiny method is compared with a set of advanced existing approaches for detecting attacks in IoT networks, using the Detection of IoT Botnet Attacks N_BaIoT dataset. Table 9 presents the performance metrics of these models, including Loss, F1-score, Recall, Precision, and Accuracy. The results indicate that the proposed method, with a Loss value of 0.0107 and an Accuracy of 99.63% in multi-class classification (8 classes), achieves performance very close to the best existing methods and even outperforms most of them. Additionally, in terms of F1-score and Recall, the proposed method shows significant improvements, achieving 99.63%, surpassing many of the compared approaches. The main reason for this improvement is the hybrid architecture, where the CNN is responsible for extracting local features from input data, while ConvNeXt-Tiny, as a modern CNN-based architecture, can extract deeper and more abstract features efficiently. The combination of these two models enables the network to effectively learn both low-level and high-level features. In the design of the proposed method, ReLU activation functions are used in the intermediate layers and Softmax in the output layer to compute more accurate probabilities for each class. This design reduces errors, increases training stability, and improves final accuracy. Figure 12 illustrates the accuracy of the proposed method compared to other existing methods. As observed, the CNN + ConvNeXt-Tiny hybrid model ranks among the most accurate approaches, with a slight difference from the highest recorded method, outperforming many well-known architectures such as CNN-LSTM, RNN, and LGBA-NN. These results demonstrate the high efficiency and effectiveness of the proposed method in detecting various attacks in resource-constrained IoT environments.

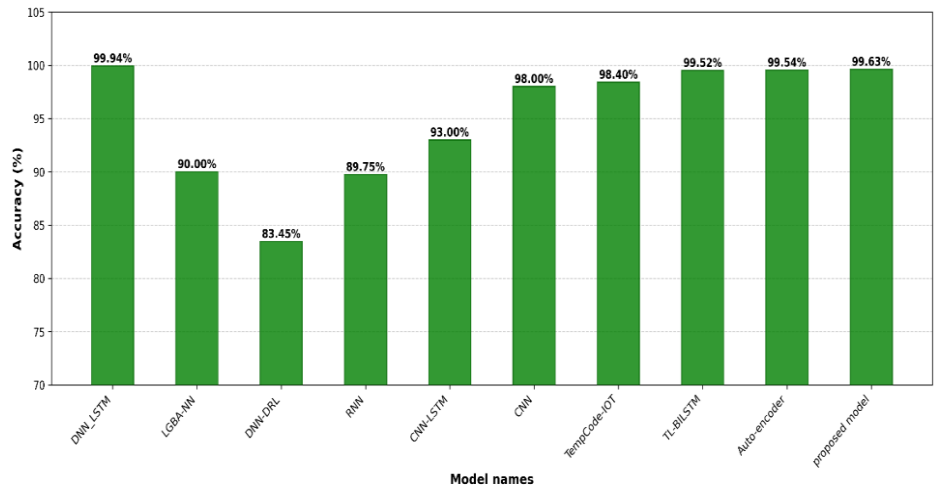

**Figure. 13:** Accuracy comparison of the proposed method with previous works

Table 9: Comparison of the proposed method with the state of artwork on "Detection of IoT botnet attacks N_BaIoT" dataset

| References | Purpose | Model name | Classification type | Side | Accuracy | Precision | Recall | F1-score | Loss |
|---|---|---|---|---|---|---|---|---|---|
| [36] | Hybrid deep learning approach for botnet attack in securing industrial IoT | DNN_LSTM | Multiclass classification (6 classes) | Client | 99.94 | 99.91 | 99.86 | 99.86 | - |
| [37] | Botnet attack detection using bio-inspired algorithm | Local–global best bat algorithm for neural network(LGBA-NN) | Multiclass classification (10 classes) | Client | 90 | 90 | 90 | 90 | 0.2 |
| [38] | Heterogeneous IoT attack detection using deep reinforcement learning | DNN-DRL | Binary classification | Client | 83.45 | - | - | - | - |
| [39] | Deep learning approach for detecting botnet attacks in IoT with heterogeneous sensors | RNN | Multiclass classification | Client | 89.75 | - | - | - | - |
| [40] | Botnet attack detection in IoT | CNN-LSTM | Multiclass classification | Client | 93 | 94 | 89 | 85 | 0.13 |
| [41] | Classification of botnet attacks in IoT | CNN | Multiclass classification (3 classes) | Client | 98 | 98 | 97 | 97.6 | - |
| [42] | Temporal codebook-based encoding for intrusion detection (flow features) | TempCode-IOT | Binary classification | Client | - | 98.4 | 99.4 | 98.9 | - |
| [19] | Effective intrusion detection framework for IoT systems | TL-BILSTM | Multiclass classification (10 classes) | Client | 99.52 | 99.54 | 99.50 | 99.52 | 0.0150 |
| [43] | Anomaly detection using transfer learning for DDoS attack | Auto-encoder | Multiclass classification(3 classes) | Client | 99.54 | - | - | - | - |
| Our proposed method | Hybrid Lightweight model for IDS in IoT | CNN+ ConvNext-Tiny | Multiclass classification (on8 classes) | Client | 99.63 | 99.63 | 99.63 | 99.39 | 0.0107 |

## 5.7. Discussion

Botnets are collections of malicious software capable of identifying vulnerable devices within a network and infecting them, thereby converting these devices into bots under the attacker's control. These bots can then be coordinated to launch large-scale cyberattacks against networks and services. To counter these threats, this study proposes a lightweight intrusion detection system based on a hybrid CNN + ConvNeXt-Tiny model as a novel method for identifying and preventing attacks in IoT environments.The proposed method leverages the Detection of IoT Botnet Attacks (N_BaIoT) dataset to extract meaningful features from network data streams and detect both known and unknown attack patterns. In this approach, the CNN component acts as a local feature extractor from network packets, while ConvNeXt-Tiny, with its modern architecture, captures higher-level and more complex features from the data. This combination leads to significant improvements in accuracy and a reduction in error rates. Experiments conducted on real data collected from IoT devices (such as IP cameras) demonstrate that the proposed method outperforms other advanced approaches. As shown in the results and graphical representations in previous sections, the method achieved 99.63% accuracy, along with high F1-score and Recall, highlighting its ability to perform multi-class classification of various attacks effectively. Overall, the hybrid CNN + ConvNeXt-Tiny model offers an efficient and computationally lightweight solution for intrusion detection in resource-constrained IoT environments. Its capability to recognize attack signatures as well as detect previously unknown patterns makes this system a reliable option for enhancing the security of IoT-based organizations and infrastructures.

## 6. Conclusion and Future Work

The Internet of Things (IoT) has experienced rapid growth in recent years, which has led to increased network traffic and, consequently, heightened security threats. To address these challenges, this paper proposed a lightweight intrusion detection system based on a hybrid CNN + ConvNeXt-Tiny model, specifically designed for IoT environments. By leveraging the deep feature extraction capabilities of CNN and the optimized, lightweight architecture of ConvNeXt-Tiny, the proposed method achieves high accuracy in detecting attacks and anomalies while reducing computational complexity. The method was trained and evaluated on the Detection of IoT Botnet Attacks (N_BaIoT) dataset and demonstrated superior performance compared to other deep learning approaches, achieving 99.63% accuracy and a validation loss of 0.0107 with a learning rate of 0.001. The results indicate that this lightweight method reduces resource consumption without sacrificing detection performance, maintaining high precision, recall, and F1-score in identifying botnet attacks. A key feature of the system is its ability to process large-scale data in real time with minimal computational overhead, making it ideal for resource-constrained devices and nodes in IoT networks. However, limitations remain, such as challenges in detecting multi-vector attacks and zero-day attacks. For future work, the proposed framework can be extended to detect other complex attacks in IoT and deployed in real-world network architectures. Additionally, evaluating the model on larger and combined datasets could enhance its ability to tackle challenging multi-class anomaly detection problems. Future plans include hyperparameter optimization to improve efficiency, deployment on low-power and decentralized hardware while preserving privacy, and development of deep learning tools for rapid and accurate detection of security threats in industrial IoT systems.